\documentclass[iop,apj]{emulateapj}
\usepackage{amsmath,amssymb,amstext}

\usepackage[breaklinks,colorlinks,citecolor=blue,linkcolor=magenta]{hyperref} 

\usepackage[all]{hypcap} %Links go to figures; breaks on deluxetables (use \capstartfalse \capstarttrue to fix it)
\usepackage[caption = false]{subfig}

\usepackage{aas_macros}
\usepackage{natbib}
\shorttitle{The Age of $\kappa$ And}
\shortauthors{Jones et al.}

\begin{document}

\defcitealias{hinkley_2013}{H13}
\defcitealias{jones_2015}{J15}
\defcitealias{david_2015}{DH15}

\title{The Age of the Directly-Imaged Planet Host Star $\kappa$ Andromedae Determined From Interferometric Observations}
\author{Jeremy Jones\altaffilmark{1}\altaffilmark{*}, R. J. White\altaffilmark{1}, S. Quinn\altaffilmark{1}, M. Ireland\altaffilmark{2}, T. Boyajian\altaffilmark{3}, G. Schaefer\altaffilmark{1}, E. K. Baines\altaffilmark{4}}

\altaffiltext{*}{Correspondence to: jones@astro.gsu.edu}
\altaffiltext{1}{Center for High Angular Resolution Astronomy and Department of Physics and Astronomy, Georgia State University, 25 Park Place, Suite 605, Atlanta, GA 30303, USA}
\altaffiltext{2}{Research School of Astronomy \& Astrophysics, Australian National University, Canberra ACT 2611, Australia}
\altaffiltext{3}{Department of Astronomy, Yale University, New Haven, CT 06511}
\altaffiltext{4}{Remote Sensing Division, Naval Research Laboratory, 4555 Overlook Avenue SW, Washington, DC 20375}

\begin{abstract}
	$\kappa$ Andromedae, an early type star that hosts a directly imaged low mass companion, is expected to be oblate due to its rapid rotational velocity ($v\sin i$ = $\sim$162 $\mathrm{km~s^{-1}}$). 
    We observed the star with the CHARA Array's optical beam combiner, PAVO, measuring its size at multiple orientations and determining its oblateness. 
    The interferometric measurements, combined with photometry and this $v\sin i$ value are used to constrain an oblate star model that yields the fundamental properties of the star and finds a rotation speed that is $\sim$85\% of the critical rate and a low inclination of $\sim$30$^\circ$. 
    Three modeled properties (the average radius, bolometric luminosity, and equatorial velocity) are compared to MESA evolution models to determine an age and mass for the star. 
    In doing so, we determine an age for the system of 47$^{+27}_{-40}$ Myr. 
    Based on this age and previous measurements of the companion's temperature, the BHAC15 evolution models imply a mass for the companion of 22$^{+8}_{-9}$ M$_\mathrm{J}$. 
\end{abstract}

\section{Introduction}
	The vast majority of exoplanets have been discovered with indirect methods such as studying the radial velocity variations induced on the host star or measuring how much light from the host star is blocked by the transiting planet \citep{winn_2015}.
	However, the spectral lines of typical early-type stars are rotationally broadened, making them not conducive to the precise radial velocity measurements necessary for planetary detection and confirmation.
    In fact, only 15 sub-stellar mass companions have been discovered around early-type stars \citep[and references therein]{hartman_2015}.
    Five of these were discovered using the transit method and the remaining ten were discovered with direct imaging.
    Accurate age estimates of stars that harbor directly imaged companions are necessary to determine the masses of the companions because these masses are all dependent on evolution models designed for low-mass objects that cool with age \citep[e.g.,][]{baraffe_2003}.
    
    The B9IVn star, $\kappa$ Andromedae A (hereafter, $\kappa$ And A; other identifiers include 19 And, HD 222439, HIP 116805, HR 8976, and T\'{e}ng Sh\'{e} \`{e}rsh\'{i}y\={\i} - The Twenty First Star of Flying Serpent) is the hottest ($T_\mathrm{eff}$ $\sim$ 11200 K) and most massive ($M$ $\sim$ 2.8 $M_\sun$) star known to host a directly imaged companion (hereafter, $\kappa$ And b), discovered by \cite{carson_2013}.
	The host star is rapidly rotating with a $v\sin i$ of $\sim$ 160 $\mathrm{km~s^{-1}}$ \citep{glebocki_2005,royer_3} and is at a distance of $51.6\pm0.5$ pc \citep{vanleeuwen_2007}.
    \cite{zuckerman_2011} consider it to be a member of the 30 Myr Columba association. 
    \cite{carson_2013} adopted this age for $\kappa$ And A and used DUSTY cooling models \citep{baraffe_2003} to determine the mass of $\kappa$ And b to be 12.8$^{+2.0}_{-1.0}$ M$_\mathrm{Jup}$.
	\citealt{hinkley_2013} (hereafter H13) estimated the age of the system to be $220\pm100$ Myr, $\sim$7 times older than the age of Columba by comparing $\log(g)$ and $T_\mathrm{eff}$ estimates to the predictions of stellar models. 
    At this age the mass of $\kappa$ And b would be 50$^{+16}_{-13}$ M$_\mathrm{Jup}$, much larger than the traditional boundary of $\sim$13 M$_\mathrm{Jup}$ between planets and brown dwarfs \citep{spiegel_2011,molliere_2012,bodenheimer_2013}. 
      
    Other studies estimate a range of ages for $\kappa$ And A. 
    \cite{bonnefoy_2014} compare the star's position on an $M_V$ vs. $B - V$ color-magnitude diagram to the predictions of the \cite{ekstrom_2012} evolution models and find an age $\lesssim$250 Myr. 
    \cite{david_2015} (hereafter \citetalias{david_2015}) use high-precision $uvby\beta$ photometry to estimate the $T_\mathrm{eff}$ and $\log(g)$ of a large sample of early-type stars, including $\kappa$ And A, and estimate ages by comparing these values to the predictions of the evolution models of \cite{bressan_2012} and \cite{ekstrom_2012}.
    With their Bayesian analysis, they find a 95\% confidence interval of 29-237 Myr for $\kappa$ And A and argue that it is not coeval with Columba. 
    Alternatively, the Bayesian analysis of \cite{brandt_2015} suggests that coevality with Columba cannot be ruled out. 
    
	To more accurately determine the properties of $\kappa$ And A, including its age, we present interferometric observations of $\kappa$ And A taken with the PAVO beam combiner on the CHARA Array. 
	Using the model described in \cite{jones_2015} (hereafter J15), we determine various fundamental parameters of $\kappa$ And A, including its radius, temperature, inclination, and luminosity; and based on comparisons with the MESA evolution model \citep{mesa1,mesa2}, determine its mass and age.
    %The luminosity determined by the model, as well as the average radius and equatorial rotation velocity, are compared to the predictions of MESA evolution models \citep{mesa1,mesa2} to determine the age and mass of the star.
    This procedure was validated using coeval members of the Ursa Major Moving Group (UMMG), showing that the MESA evolution models are appropriate for dating rapidly rotating stars by finding coeval ages between rapidly and slowly rotating members of the UMMG and by estimating an age for the group in agreement with the admittedly large range of age estimates for the group. 
    With an age for the $\kappa$ And system, we estimate a mass for the companion by using the BHAC15 evolution models \citep{baraffe_2015}.

\section{Observations and Data Reduction}
\subsection{Visibilities}
	Observations of $\kappa$ And A were made using the PAVO (Precision Astronomical Visible Observations) beam combiner on the CHARA (Center for High Angular Resolution Astronomy) Array \citep{pavo,chara2}.
    The CHARA Array is an optical interferometer made up of six 1-m telescopes arranged in a Y-shaped configuration with a maximum baseline of 331 m.
    Each telescope is named with a letter designating its arm (``S''-south, ``E''-east, ``W''-west) and a number designating its place on the arm (``1"-outer, ``2"-inner).
    PAVO was used in its two-telescope mode and produces 23 spectrally dispersed squared-visibility measurements for each observation over a wavelength range of 0.65-0.79 $\mu$m.
    In total, we made 24 observations yielding 552 spectrally-dispersed squared-visibility measurements over four nights using five different baselines in order to measure its oblateness.
    
    We observe two different calibrator stars (HD 222304 and HD 220885) shortly before and after (within $\sim$30 minutes) our observations of $\kappa$ And A and by doing so, we can account for how the atmosphere dampens the measured visibilities of the target star \citep{boden_2007,roddier_1981}.
    We predict that these calibrator stars have small angular diameters ($<0.27$ mas) based on fitting photometric energy distributions to measured photometry.
    %We used the reduction pipeline of \cite{pavo} to reduce and calibrate the data.
    We reduce and calibrate the data with the reduction pipeline of \cite{pavo}.
    Table \ref{tab:log} lists the dates observations were made, how many observations were made, the baselines used, and the calibrator used. 
    
\subsection{Photometry}

	We take advantage of the ample photometric observations of $\kappa$ And A that have been made over the years, using photometry from the following sources - Johnson $UBV$ from \cite{mermilliod_1991}; Str{\"o}mgren $uvby$ from \cite{hauck_1997}; Johnson $JK$ from \cite{selby_1988}; and UV photometry with wavelengths ranging from 1500 {\AA} to 3300 {\AA} from \cite{thompson_1978} and \cite{wesselius_1982}. 
    IUE spectrophotometry \citep{iue} exists for $\kappa$ And A that we do not use, but matches to our model spectral energy distribution (SED) and the broadband UV photometry that we use. 
    Following arguments from \citetalias{jones_2015}, we adopt an uncertainty of 0.03 mag for all photometric values. 

\begin{table*}
	\begin{center}
	\caption{Observing Log. \label{tab:log}}
	\begin{tabular}{cccccc}
	\tableline\tableline
	Cal HD 		& Cal Diameter (mas)  	& Baseline 	& \# Observations 	& \# visibilities 	& Date \\
	\tableline
	222304 		& 0.263 $\pm$ 0.026 	& S2-E2 	&  4			& 92 				& 12/21/2012 \\
	220885 		& 0.230 $\pm$ 0.023 	& S2-E2 	&  4			& 92	 			& 12/21/2012 \\
	222304	 	& 0.263 $\pm$ 0.026 	& W1-E1 	&  1			& 23 				& 8/2/2013 \\
	220885 		& 0.230 $\pm$ 0.023 	& S1-E1 	&  2			& 46		 		& 8/2/2013 \\
	220885	 	& 0.230 $\pm$ 0.023 	& S1-E1 	&  3			& 69 				& 8/3/2013 \\
	220885	 	& 0.230 $\pm$ 0.023 	& W1-S1 	&  3			& 69 				& 8/3/2013 \\
	222304	 	& 0.263 $\pm$ 0.026 	& W1-S1 	&  3			& 69 				& 8/3/2013 \\
	220885	 	& 0.230 $\pm$ 0.023 	& E1-W2 	&  4			& 92 				& 8/5/2013 \\
	 \tableline\tableline
	\end{tabular}
	\end{center}
\end{table*}

\section{Modeling of Stellar Properties}
\label{sec:model}
\subsection{Oblate Star Model}
	Because of $\kappa$ And A's rapid rotation ($v\sin i=161.6\pm22.2$ $\mathrm{km~s^{-1}}$, \citep{glebocki_2005,royer_3}), the limb-darkened disk traditionally used to model interferometric data is insufficient. 
    Rapid rotation causes a star to have a radius at the equator larger than its radius at the pole.
    The ratio between the equatorial and polar radii can be as high as 1.5 when the star is rotating at its breakup velocity \citep{vanbelle_2012}. 
    The thicker equatorial bulge of a rapid rotator results in the equator being both cooler and fainter than the pole. 
    This effect, known as gravity darkening, is correlated with the local surface gravity \citep{von_zeipel_1924a,von_zeipel_1924b}.
   
    We account for both the oblateness and gravity darkening of $\kappa$ And A by using the model of \citetalias{jones_2015}, which compares observed photometry and interferometric visibilities to values generated by a model star that incorporates the effects of solid-body rotation, known as a `Roche model' \citep{vanbelle_2012,roche_1873}. 
    The model photometry are calculated by integrating ATLAS model SEDs \citep{castelli_2004} over the visible surface of the star, convolving the integrated SED with the appropriate filter bandpasses, and converting the resulting fluxes into magnitudes. 
    To calculate model visibilities, we generate an image of the model at the observed bandpasses. 
    The model visibilities are calculated by taking the Fourier transform of this image and sampling the transform at the observed spatial frequencies. 
    
    The model and parameters calculated by the model are described in detail in \citetalias{jones_2015}, but we note three slight differences here. 
    %Because of the length constraints of a letter, we do not describe the model here, but we do discuss differences between the analysis presented here and that used on the UMMG stars in \citetalias{jones_2015}. 
    One such difference is that we use ATLAS model SEDs for this work rather than the PHOENIX model SEDs used in \citetalias{jones_2015} \citep{phoenix}, since they extend to effective temperatures hotter than 12000 K. 
    %This is because the PHOENIX model SEDs are limited to temperatures less than $T_\mathrm{eff} = 12000$ K, less than the polar temperature we model for $\kappa$ And A. 
    Another difference is that we only use the gravity darkening law of \cite{ELR_2011}, because the data are not sensitive to differences in gravity darkening laws and this law is supported by previous interferometric observations. 
    
    The final difference is in how uncertainties are calculated. 
    Under the assumption that the uncertainties in the free parameters are Gaussian and that the model parameters are linear, we use the following prescription to determine uncertainties in the free parameters: 
	Because the $\chi^2$ values determined by the models are larger than 1, for each data set (photometry and visibilities), we scale the $\chi^2$ (both reduced and unreduced) such that the reduced $\chi^2$ is 1. 
    The free parameters are then varied individually until the scaled, unreduced $\chi^2$ increases by 1. 
    This gives two sets of uncertainties for the free parameters - one for the photometry and one for the visibilities, with the exception of the position angle, which is only probed by the visibilities. 
    The final uncertainty in each free parameter is determined by adding the two uncertainties in quadrature under the assumption that the visibilities and photometry are independent. 
    The uncertainty in the position angle is determined only by comparison with the visibilities. 
    These uncertainties are then propagated to determine the uncertainties in the derived parameters.
    We caution the reader that these uncertainties are statistical and do not account for systematic uncertainties such as errors in the model spectra, gravity darkening law, etc. 
    The coevality of oblate and non-oblate A-stars in the UMMG, determined using this model \citepalias{jones_2015}, suggests that these systematic uncertainties do not dominate the errors. 
    
\begin{figure*}
	\subfloat[\label{fig:solar_ELR_visanddiff}]{\includegraphics[height =2.5in]{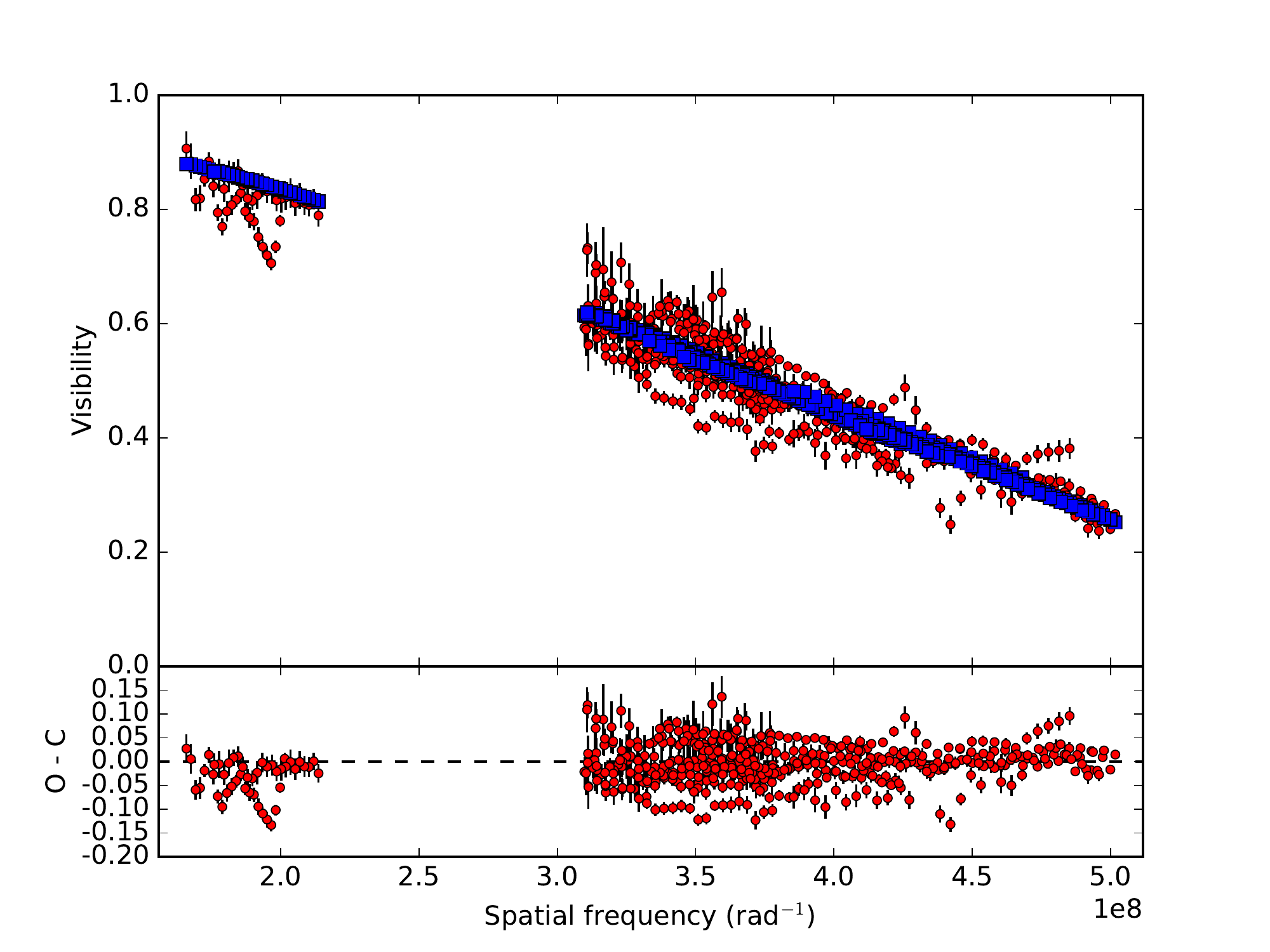}} \\
	\subfloat[\label{fig:solar_ELR_photanddiff}]{\includegraphics[height =2.5in]{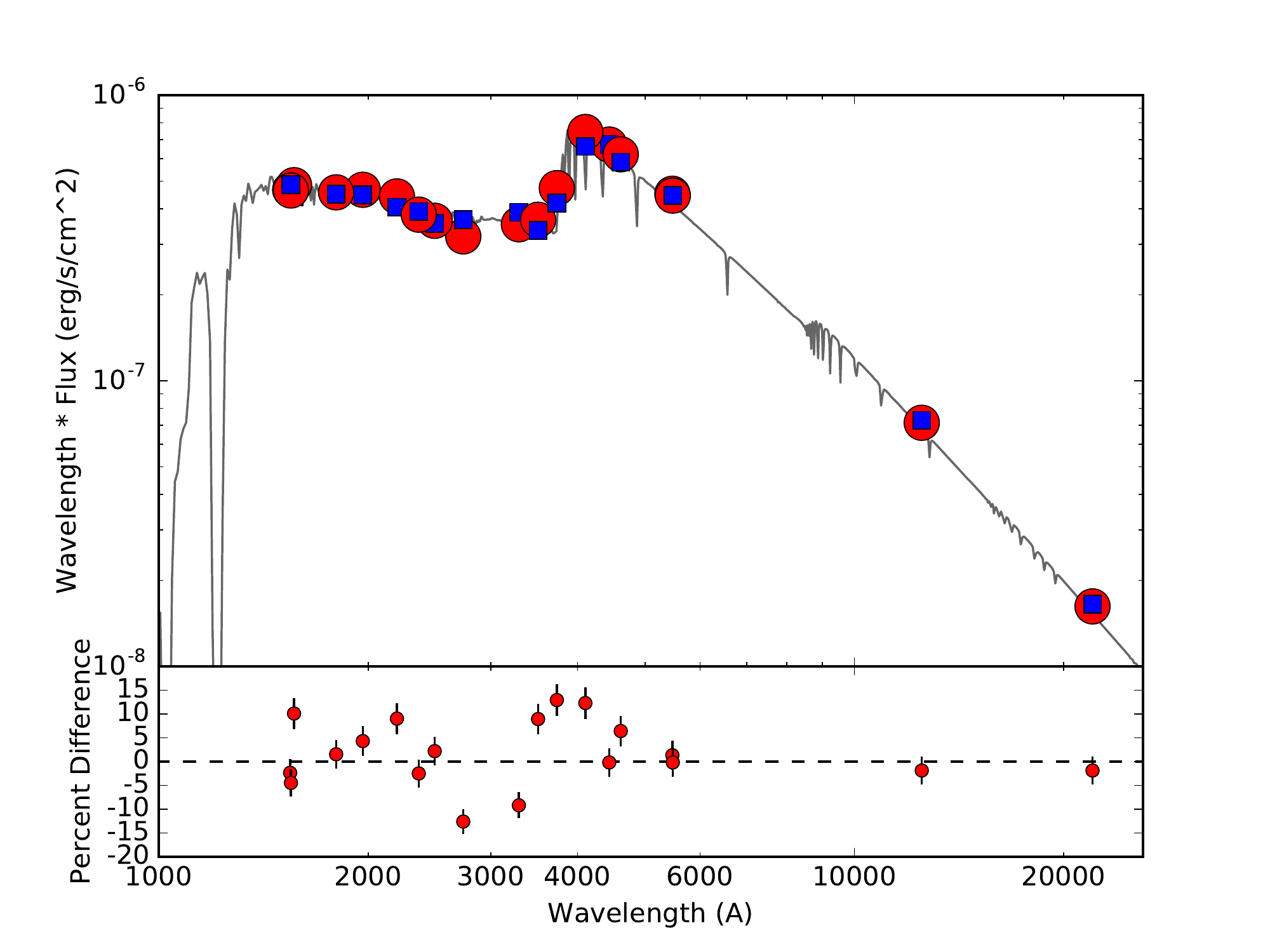}} \\
	\subfloat[\label{fig:solar_ELR_ellplot}]{\includegraphics[height =2.5in]{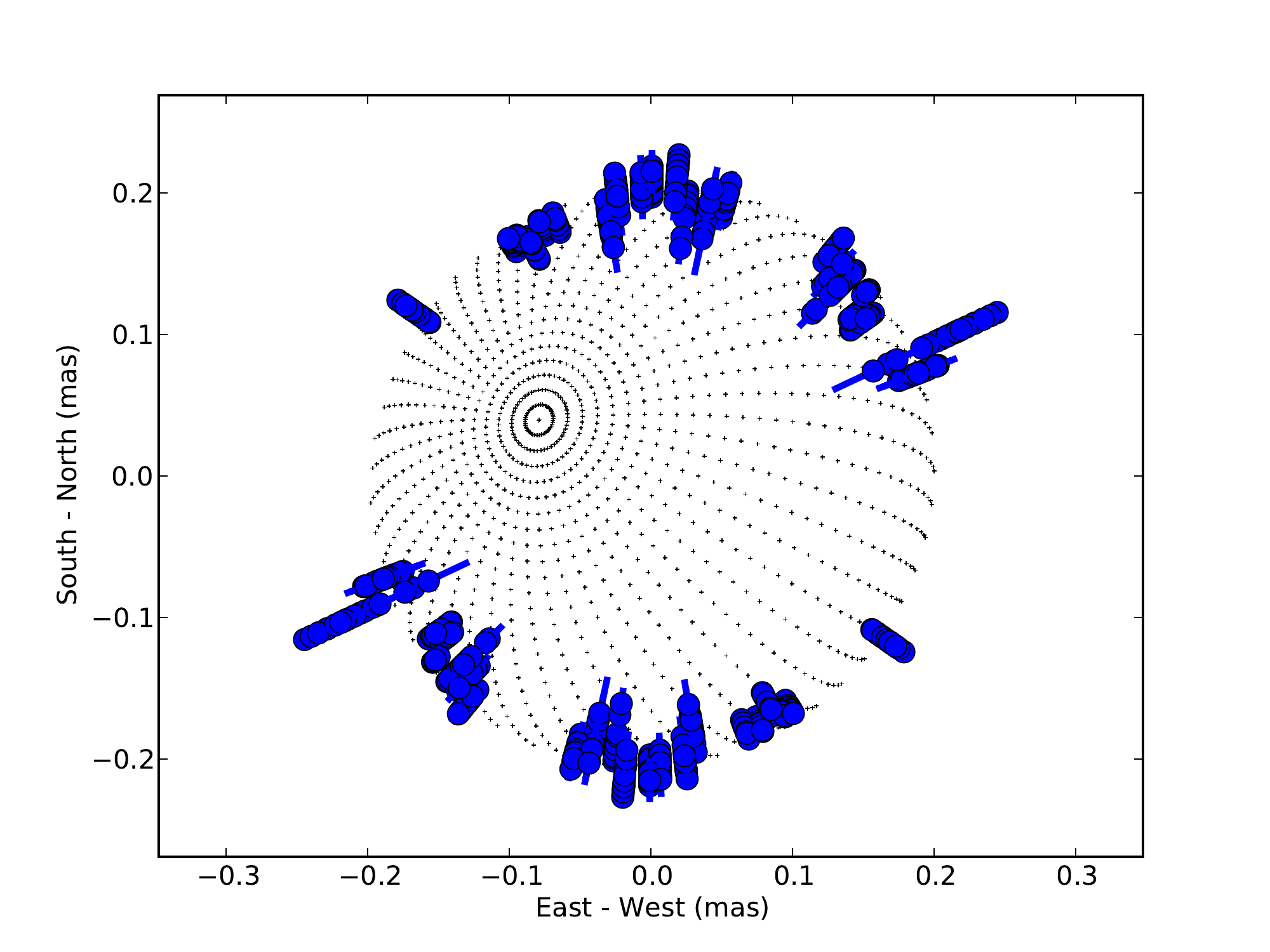}} \\
	\caption{Top: Observed (red circles) and best-fit model visibilities (blue squares) vs. spatial frequencies for the solar metallicity model. Middle: Observed (red circles) and best-fit model (blue squares) photometric fluxes vs. wavelength for the solar metallicity model. The modeled SED is shown in gray. Bottom: The photosphere of the best fitting model of $\kappa$ And A. The black points represent a grid of colatitudes and longitudes on the near side of the model. The blue circles represent a radius fitted to each individual visibility at the appropriate baseline orientation observed. The data are duplicated at 180$^\circ$ orientation.}
	\label{fig:solar_ELR}
\end{figure*}

    Figure \ref{fig:solar_ELR} illustrates the best fitting model by showing the modeled visibilities and photometry as well as the modeled photosphere overlaid with approximate radius measurements at various orientations. 
    %The best-fit modeled properties and their derived properties are listed in Table \ref{tab:results}.
    Using four different metallicities (justified below), the best-fit modeled properties are listed in rows 3 - 7 of Table \ref{tab:results}, and the properties derived from these are in rows 8 - 20 of Table \ref{tab:results}.

\subsection{Stellar Evolution Models}
\label{sec:evo_mod_desc}
	We take the average radius ($R_\mathrm{avg}$), total bolometric luminosity ($L_\mathrm{bol}$), and equatorial rotation velocity ($V_\mathrm{e}$) shown in Table \ref{tab:results} and use MESA evolution models \citep{mesa1,mesa2} to determine the age and mass of $\kappa$ And A by comparing the modeled values to MESA's predictions for given masses, ages, and initial rotation rates. 
   MESA models are used because they can account for the rapid rotation of $\kappa$ And A. 
    The uncertainties in the mass and age are based on propagated uncertainties in stellar properties \citepalias{jones_2015}. 
    %determined using the same method used in \citetalias{jones_2015}.}
    
\begin{figure*}
	\includegraphics[height =5in]{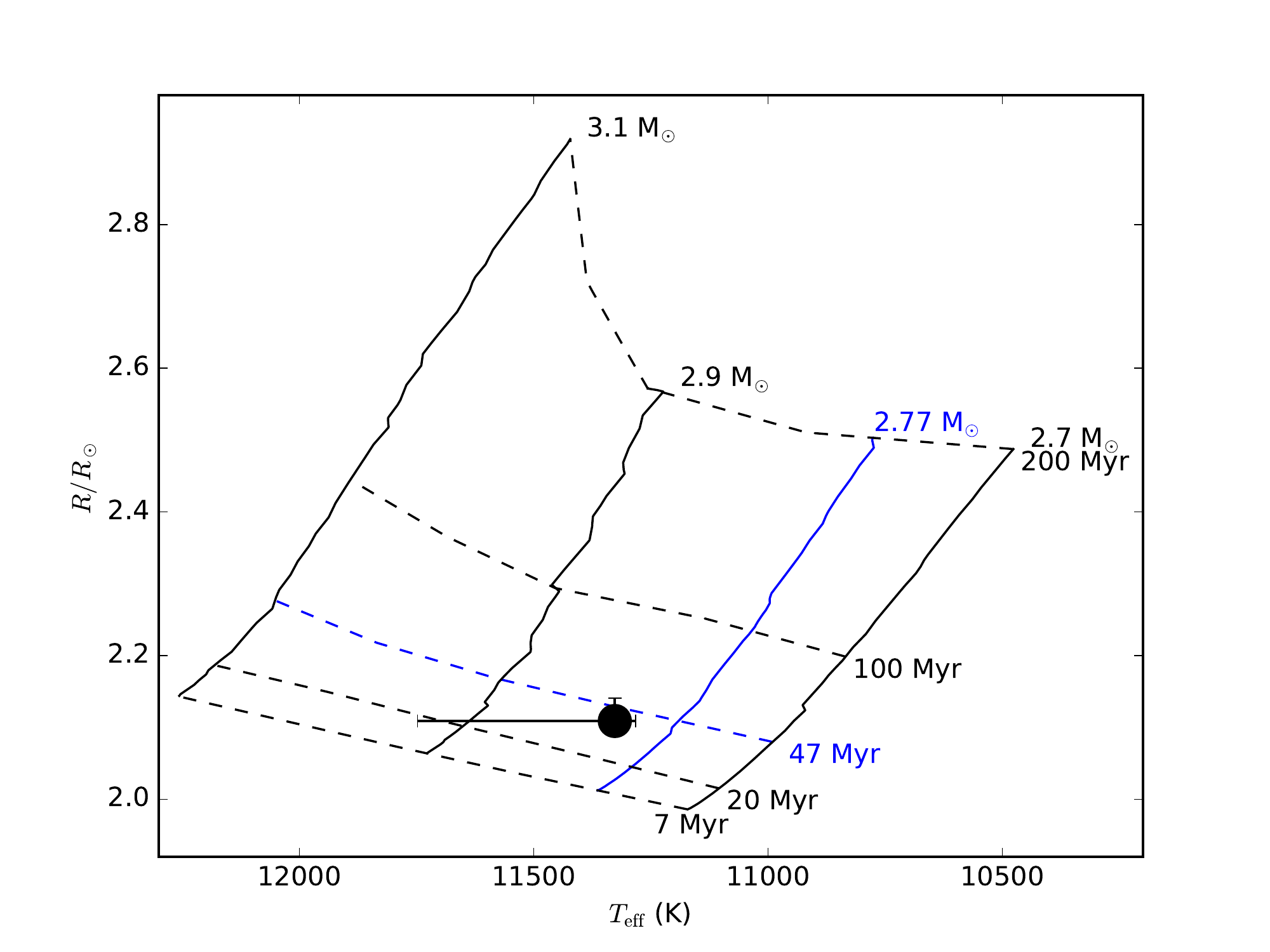}
    \caption{
      The solid lines show the evolution in radius and effective temperature according to the mass tracks of the MESA evolution models for masses ranging from 2.7 to 3.1 M$_\sun$.
      The dashed lines are isochrones showing the radius and effective temperatures of stars with this range of masses at ages ranging from 7 to 200 Myr. 
      Both the mass tracks and isochrones were calculated for solar metallicity and interpolated to the modeled rotation velocity of the star.
    }
    \label{fig:RvT}
\end{figure*}

    One systematic source of uncertainty that is difficult to account for in this analysis is the metallicity of the evolution model.
    There are several reasons to suspect that the subsolar surface abundance of $\kappa$ And A (e.g. [M/H] = $-0.32\pm0.15$; \cite{wu_singh_2011}) does not trace its internal abundance.
    First, the surface abundances of A- and B-stars within populations believed to be chemically homogeneous span a broad range.
    Moreover, there is evidence that photospheric abundances are anti-correlated with projected rotational velocity ($v\sin i$), becoming distinctively subsolar (e.g., $\lesssim-0.30$) when projected rotational velocities exceed $\sim$150 km/s \citep[e.g.,][]{takeda_1997,varenne_1999}. 
    Thus, there is reason to suspect that the internal abundance of $\kappa$ And A is more metal rich than is observed in its photosphere.  
    Finally, as emphasized by \citetalias{hinkley_2013}, the Galaxy has not recently produced many stars that are this metal poor.  
    To quantify this, we consider the sample of open clusters with metallicty measurements assembled in \cite{chen_2003}.  
    These 77 clusters have a mean metallicity of 0.00 dex and a standard deviation of 0.14 dex; the most metal poor cluster among them has a metallicty of $-$0.34 dex. 
    Given these consideration, we adopt a solar metallicity ([M/H]=0.00 dex, Z=0.0153, \citealt{caffau_2011}) for $\kappa$ And A, with an uncertainty of 0.14 dex.
    Nevertheless, we also consider a metallicity of [M/H]=$-$0.28 dex as a 2$\sigma$ extremum in our analysis.
    Figure \ref{fig:RvT} shows the average radius and temperature of $\kappa$ And A overlaid with mass tracks and isochrones from the MESA evolution models for solar metallicity which have been interpolated to the modeled rotational velocity.
    
\section{Results and Discussion}
\subsection{The Properties of $\kappa$ And A}
\label{res:kap_and_a}
	We use the model discussed in Section \ref{sec:model} to determine the age of $\kappa$ And A for four different internal metallicities ([M/H]=$+$0.14, 0.0, $-$0.14, and $-$0.28) corresponding to the $+$1-, 0-, $-$1-, and $-$2$\sigma$ uncertainties in [M/H], respectively. 
    For the solar metallicity model, we find a radius for the host star ranging from $2.303^{+0.039}_{-0.016}$ R$_\sun$ at the equator to $1.959^{+0.033}_{-0.028}$ R$_\sun$ at the pole with an average of $2.109^{+0.032}_{-0.018}$ R$_\sun$. 
    This oblateness is, in part, due to an equatorial velocity of $283.8^{+13.4}_{-16.1}$ $\mathrm{km~s^{-1}}$, which corresponds to an angular rotation rate relative to the critical rate, $\omega$, of $0.854^{+0.021}_{-0.028}$ and which with the modeled inclination of $30.1^{+3.1}_{-4.8}$ $^{\circ}$ gives a modeled $v\sin i$ of $142.2^{+13.1}_{-21.1}$ $\mathrm{km~s^{-1}}$.
    Our modeled effective temperature ranges from $12050^{+448}_{-39}$ K at the pole to $10342^{+384}_{-138}$ K at the equator with an average of $11327^{+421}_{-44}$ K, and together with the modeled radius profile, yield a total luminosity of $62.60^{+9.83}_{-2.23}$ L$_\sun$ and apparent luminosity of $72.01^{+11.17}_{-1.50}$ L$_\sun$.
    We model an average surface gravity ($\log(g_\mathrm{avg})$) of $4.174^{+0.019}_{-0.012}$ dex, which is only slightly larger than previous measurements of the star's $\log(g)$ ranging from 3.8 to 4.1 dex \citep{bonnefoy_2014,fitzpatrick_2005,wu_singh_2011}. 
    %One way to possibly address this discrepancy is to adapt our model to compute spectral line profiles and compare to observed spectral lines.
    
	The age and mass we determine using the best fitting model with a solar metallicity are $47^{+14}_{-21}$ Myr and $2.768^{+0.121}_{-0.013}$ M$_\sun$, respectively.
    This young age is due, in large part, to the low inclination ($\sim$30$^\circ$) and large rotation velocity ($\sim$85\% critical) which implies that the apparent luminosity is brighter than the total luminosity because of the effects of gravity darkening and which also changes where the zero-age main sequence (ZAMS) lies on the HR diagram. 
    
    Most of our modeled parameters show broad agreement between the four different internal metallicities tested, however the age and the mass show a significant correlation with metallicity (e.g., a lower metallicity corresponds to an older age and a lower mass).
    Given how strongly the internal metallicity affects the modeled mass and age of the host star, we adopt the ages and masses determined at the 1$\sigma$ uncertainties in the metallicity as the bounds to our final uncertainties in the age and mass.
    The supersolar metallicity model ([M/H]=$+$0.14) has a radius and luminosity below the ZAMS, so we adopt the age of the ZAMS, $\sim$7 Myr, as the lower bound of the uncertainty in the age.
    Given the trend of decreasing mass of $\sim$0.1 $M_\sun$ for every 1$\sigma$ decrease in metallicity, we adopt an upper bound of the uncertainty in our mass to be 0.1 $M_\sun$
    Thus, our final estimate of the age and mass of $\kappa$ And A is $47^{+27}_{-40}$ Myr and $2.768^{+0.1}_{-0.109}$ M$_\sun$, respectively.

	We note that a more recent age estimate of the Columba association by \cite{bell_2015} finds it to be $42^{+6}_{-4}$ Myr, which is in excellent agreement with our age estimate for $\kappa$ And A.
    Despite its outlying Galactic Y position with respect to Columba (2.7$\sigma$, \citetalias{hinkley_2013}), the agreement in age suggests that its kinematic association with young nearby groups should be reconsidered. 

\subsection{A Comparison to Previous Age Estimates}
	\citetalias{hinkley_2013} use a variety of methods to estimate the age of $\kappa$ And A, finding ages ranging from $\sim$50-400 Myr.
    Their adopted age of 220 $\pm$ 100 Myr is based on a comparison between the predictions of the Geneva evolution models \citep{ekstrom_2012} which account for a rotation rate of $\omega$=0.4 and the $\log(g)$ (4.10 dex) and $T_\mathrm{eff}$ (11366 K) measured by \cite{fitzpatrick_2005}.
    This age estimate is significantly older than both the traditionally adopted age of the Columba association (30 Myr) and our estimate ($47^{+27}_{-40}$ Myr).
    \citetalias{hinkley_2013} do note that such a young age is possible if the host star is rapidly rotating ($V_\mathrm{e}/V_\mathrm{crit}\simeq0.95$) with an very low orientation ($\simeq22^\circ$), which is what we have found with this work.

	\citetalias{david_2015} use Str{\"o}mgren photometry of \cite{hauck_1997} to determine a $\log(g)$ of 4.35 $\pm$ 0.14 dex and $T_\mathrm{eff}$ of 11903 $\pm$ 405 K.
    From this, they interpolate between the isochrones generated by the evolution models of \cite{bressan_2012} and \cite{ekstrom_2012} to estimate an age of 16 Myr.
    Superseding this interpolated estimate, they use a more thorough Bayesian approach and find a 95\% confidence interval of 29-237 Myr with a median age of 150 Myr.

	In an attempt to determine how much the choice of evolution model affects the estimated age, we compare the $\log(g)$ and $T_\mathrm{eff}$ values used by both \citetalias{hinkley_2013} and \citetalias{david_2015} to the MESA evolution models used here. 
    We estimate an age of 185 Myr and 13 Myr using the $\log(g)$ and $T_\mathrm{eff}$ values used by \citetalias{hinkley_2013} and \citetalias{david_2015}, respectively.
    These estimates are lower than the estimates made by these two studies by $\sim$20\%, which is smaller than the uncertainties in the age estimates.
    %Finally, though it is beyond the scope of this letter, given the differences between the interpolated age and the Bayesian age estimated by \citetalias{david_2015}, it may be worth adapting our model to use a Bayesian method for age estimation.}

\subsection{The Mass of $\kappa$ And b}
\label{res:kap_and_b}
\begin{figure*}
	\includegraphics[height = 5in]{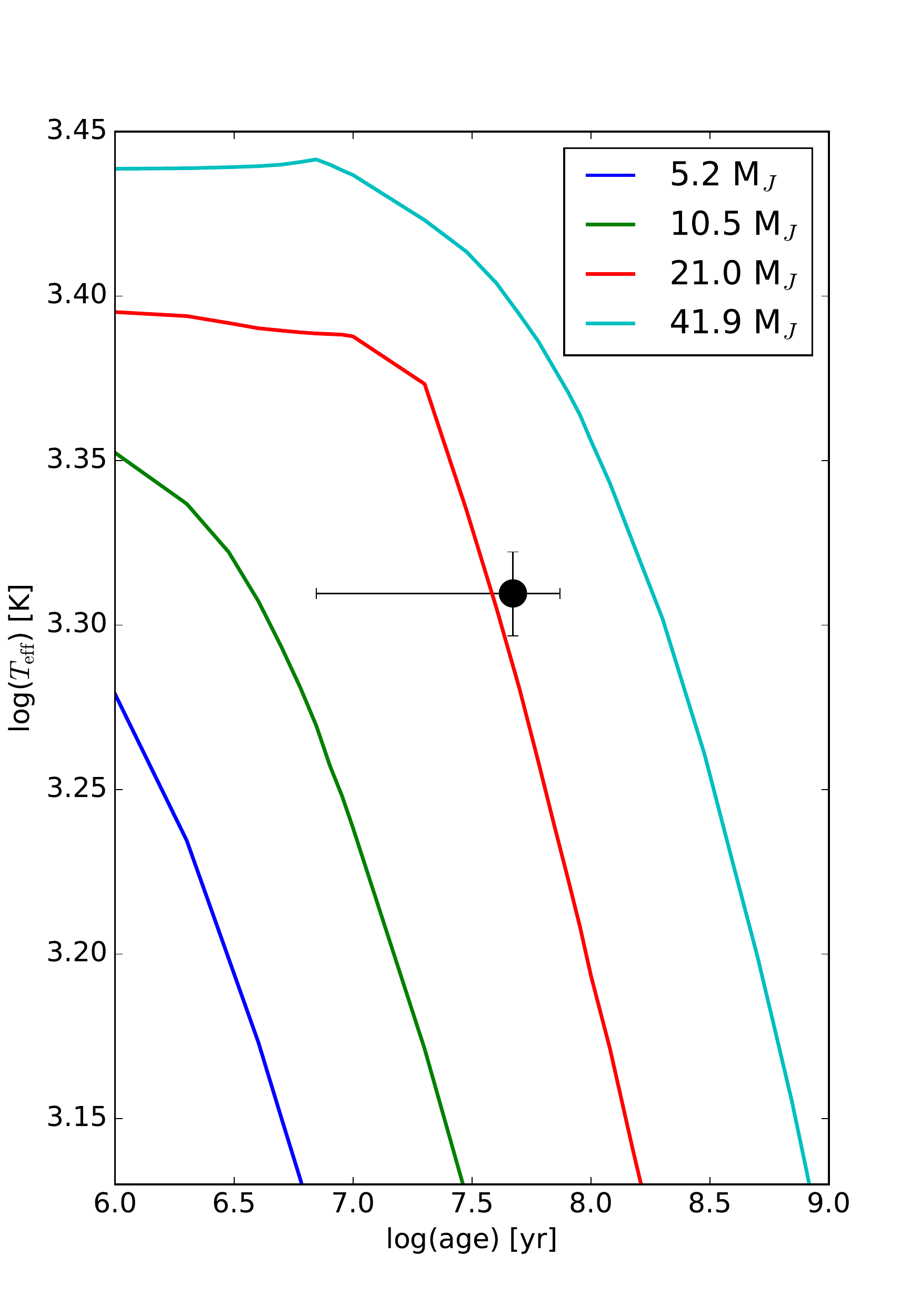}
    \caption{
      The solid lines show how the BHAC15 evolution models predict substellar objects cool over time for masses ranging from 5.2 to 41.9 M$_\mathrm{J}$. 
      The black point shows the effective temperature of $\kappa$ And b (2040 $\pm$ 60 K; \citetalias{hinkley_2013}) and its age ($47^{+27}_{-40}$ Myr; This work).}
    \label{fig:cool_plot}
\end{figure*}

    In order to determine the mass of $\kappa$ And b, we compare our age estimate for the host star and the spectroscopically determined effective temperature of the companion (2040 $\pm$ 60 K; \citetalias{hinkley_2013}) to the predictions of the updated BHAC15 models of \cite{baraffe_2015}.
    Uncertainties in the companion mass are determined by using this method to calculate the mass corresponding to the four points representing the 1$\sigma$ uncertainties in the age and effective temperature of the companion. 
    With this technique, we find a mass of $22^{+8}_{-9}$ M$_\mathrm{J}$ with the uncertainties dominated by the uncertainty in the age which is dominated by the uncertainty in the metallicity.
    Figure \ref{fig:cool_plot} shows the effective temperature of $\kappa$ And b from \citetalias{hinkley_2013} and our final estimate for the age of the system along with the cooling tracks of the BHAC15 models. 
    
\section{Summary}
	We present new PAVO/CHARA interferometric observations of $\kappa$ And A. 
    Using these observations, the star's photometry, and its $v\sin i$, we constrain an oblate star model from which we calculate various fundamental parameters. 
    These parameters include the star's luminosity, radius profile, and equatorial rotation velocity which are compared to the predictions of the MESA evolution models in order to estimate an age and mass for the star. 
    Four internal metallicities ([M/H]=$+$0.14, 0.0, $-$0.14, and $-$0.28) are tested and we find that metal-rich models yield a higher mass and younger age more metal-poor models. 

	Because the internal metallicity of the star is expected to be solar ([M/H]=0.00$\pm$0.14), we adopt the solar metallicity model with the uncertainties in our final age and mass governed by the uncertainty in the metallicity. 
    With this model, we determine an age of $47^{+27}_{-40}$ Myr for the system and a mass of $2.768^{+0.1}_{-0.109}$ M$_\sun$ for $\kappa$ And A. 
    Based on this age, the effective temperature of the companion, and the BHAC15 evolution models, we determine a mass of $\kappa$ And b of $22^{+8}_{-9}$ M$_\mathrm{Jup}$.
    
\acknowledgements
	J.J. thanks Sasha Hinkley and Josh Schlieder for their helpful comments. The authors would also like to thank the CHARA team for their hard work maintaining and operating the Array, namely Theo ten Brummelaar, Hal McAlister, Steve T. Ridgway, Judit Sturmann, Laszlo Sturmann, Nils Turner, Chris Farrington, Nic Scott, and Norm Vargas. J.J. and R.J.W. aknowledge support from the NSF AAG grants numbered 1009643 and 1517762.

\begin{table*}
\begin{center}
	\caption{Model Results. \label{tab:results}}
	\begin{tabular}{ccccc}
		\tableline\tableline
        \multicolumn{5}{c}{Properties of $\kappa$ And A} \\
		\tableline
		Internal [M/H] & +0.14 & 0.00$^a$ & -0.14 & -0.28 \\
        Internal Z & 0.0211 & 0.0153 & 0.0111 & 0.0080 \\
        \tableline
        \multicolumn{5}{c}{Modeled Properties} \\
		\tableline
		Equatorial Radius, $R_\mathrm{e}$ (R$_{\sun}$) & $2.331^{+0.068}_{-0.011}$ & $2.303^{+0.039}_{-0.016}$ & $2.326^{+0.029}_{-0.023}$ & $2.366^{+0.023}_{-0.027}$ \\
		Equatorial Velocity, $V_\mathrm{e}$ ($\mathrm{km~s^{-1}}$) & $354.8^{+7.3}_{-35.4}$ & $283.8^{+13.4}_{-16.1}$ & $322.5^{+19.2}_{-13.3}$ & $376.6^{+14.4}_{-11.5}$ \\
		Stellar Inclination, $i$ ($^\circ$) & $26.3^{+1.0}_{-7.9}$ & $30.1^{+3.1}_{-4.8}$ & $27.0^{+2.2}_{-3.2}$ & $25.9^{+2.3}_{-1.9}$ \\
		Polar Temperature, $T_\mathrm{p}$ (K) & $12195^{+144}_{-177}$ & $12050^{+448}_{-39}$ & $12167^{+314}_{-46}$ & $12348^{+47}_{-322}$ \\
		Polar Position Angle, $\psi$ ($^{\circ}$) & $69.6^{+3.2}_{-0.9}$ & $63.4^{+5.2}_{-1.0}$ & $69.3^{+0.5}_{-2.7}$ & $72.8^{+0.9}_{-1.2}$ \\
		\tableline
        \multicolumn{5}{c}{Properties Derived from Oblate Star Model} \\
		\tableline
		Gravity Darkening, $\beta$ & $0.181^{+0.011}_{-0.002}$ & $0.202^{+0.004}_{-0.004}$ & $0.188^{+0.004}_{-0.006}$ & $0.166^{+0.004}_{-0.006}$ \\
		Angular Rotation Rate, $\omega$ & $0.947^{+0.007}_{-0.039}$ & $0.854^{+0.021}_{-0.028}$ & $0.921^{+0.021}_{-0.017}$ & $0.978^{+0.008}_{-0.008}$ \\
		Polar Radius, $R_\mathrm{p}$ (R$_{\sun}$) & $1.827^{+0.078}_{-0.016}$ & $1.959^{+0.033}_{-0.028}$ & $1.878^{+0.030}_{-0.043}$ & $1.761^{+0.028}_{-0.034}$ \\
		$^b$Average Radius, $R_\mathrm{avg}$ (R$_{\sun}$) & $2.026^{+0.056}_{-0.012}$ & $2.109^{+0.032}_{-0.018}$ & $2.062^{+0.022}_{-0.031}$ & $1.983^{+0.022}_{-0.029}$ \\
		$^c$Average Angular Diameter, $\theta_\mathrm{avg}$ (mas) & $0.365^{+0.010}_{-0.002}$ & $0.380^{+0.006}_{-0.003}$ & $0.371^{+0.004}_{-0.006}$ & $0.357^{+0.004}_{-0.005}$ \\
		Equatorial Temperature, $T_\mathrm{e}$ (K) & $9662^{+414}_{-140}$ & $10342^{+384}_{-138}$ & $9933^{+256}_{-231}$ & $9222^{+175}_{-240}$ \\
		$^b$Average Temperature, $T_\mathrm{avg}$ (K) & $11250^{+133}_{-163}$ & $11327^{+421}_{-44}$ & $11290^{+291}_{-53}$ & $11307^{+43}_{-295}$ \\
		Polar Surface Gravity, $\log(g_\mathrm{p})$ (cgs) & $4.373^{+0.008}_{-0.036}$ & $4.296^{+0.019}_{-0.012}$ & $4.315^{+0.018}_{-0.013}$ & $4.355^{+0.017}_{-0.014}$ \\
		$^b$Average Surface Gravity, $\log(g_\mathrm{avg})$ (cgs) & $4.207^{+0.004}_{-0.022}$ & $4.174^{+0.019}_{-0.012}$ & $4.164^{+0.014}_{-0.009}$ & $4.169^{+0.011}_{-0.014}$ \\
		Equatorial Surface Gravity, $\log(g_\mathrm{e})$ (cgs) & $3.813^{+0.091}_{-0.041}$ & $3.968^{+0.028}_{-0.025}$ & $3.848^{+0.032}_{-0.054}$ & $3.593^{+0.054}_{-0.082}$ \\
		$v\sin i$ ($\mathrm{km~s^{-1}}$) & $157.4^{+5.8}_{-44.9}$ & $142.2^{+13.1}_{-21.1}$ & $146.2^{+11.0}_{-16.2}$ & $164.7^{+13.4}_{-11.5}$ \\
		Total Luminosity, $L_\mathrm{tot}$ (L$_{\sun}$) & $55.21^{+5.67}_{-3.14}$ & $62.60^{+9.83}_{-2.23}$ & $58.35^{+6.26}_{-3.08}$ & $53.50^{+1.73}_{-5.37}$ \\
		Apparent Luminosity, $L_\mathrm{app}$ (L$_{\sun}$) & $71.17^{+3.72}_{-3.99}$ & $72.01^{+11.17}_{-1.50}$ & $72.49^{+7.67}_{-2.04}$ & $72.99^{+1.24}_{-7.22}$ \\
		\tableline
		Visibility $\chi^2$ & 12.99 & 13.23 & 13.01 & 12.85 \\
		Photometry $\chi^2$ & 9.68 & 8.92 & 8.74 & 8.75 \\
		Total $\chi^2$ & 22.67 & 22.15 & 21.75 & 21.60 \\
		\tableline
        \multicolumn{5}{c}{Properties Derived from MESA Evolution Models} \\
		\tableline
		Age (Myr) & Below ZAMS & $47^{+14}_{-21}$ & $74^{+21}_{-28}$ & $82^{+29}_{-28}$ \\
		Mass (M$_{\sun}$) & Below ZAMS & $2.768^{+0.121}_{-0.013}$ & $2.659^{+0.087}_{-0.014}$ & $2.558^{+0.013}_{-0.084}$ \\
		\tableline
        \multicolumn{5}{c}{Properties of $\kappa$ And b} \\
		\tableline
        $T_\mathrm{eff}$ (K) \citepalias{hinkley_2013} & \multicolumn{4}{c}{$2040\pm60$} \\
        Mass (M$_\mathrm{Jup}$) & N/A & $22^{+6}_{-7}$ & $30^{+3}_{-8}$ & $31^{+4}_{-5}$ \\
		\tableline
        \multicolumn{5}{c}{Adopted System Properties using [M/H] = 0.00} \\
		\tableline
        Age (Myr) & \multicolumn{4}{c}{$47^{+27}_{-40}$} \\
        Mass of A (M$_{\sun}$) & \multicolumn{4}{c}{$2.768^{+0.1}_{-0.109}$} \\
        Mass of b (M$_\mathrm{J}$) & \multicolumn{4}{c}{$22^{+8}_{-9}$} \\
		\tableline
	\end{tabular}
    %\tablenotetext{1}{The modeled average radius and luminosity are below the ZAMS for the evolution model corresponding to the supersolar metallicity and the modeled equatorial velocity. Therefore, we adopt for the age for this model the age of the ZAMS and we adopt a mass following the trend of $\sim$ $-$0.1 $M_\sun$ for each 0.14 dex in [M/H].}
    \tablenotetext{1}{We adopt as our final results those from the solar metallicity models.}
    \tablenotetext{2}{The average quantities presented here are averaged across the entire surface of the model star.}
    \tablenotetext{3}{The average angular diameter is determined using the average radius and the distance.}
    \tablenotetext{4}{From \citetalias{hinkley_2013}}
\end{center}
\end{table*}

%\keywords{keywords}
\maketitle

\bibliography{bibliography}{}

\end{document}